# Efficient Photocatalytic H$_2$ Evolution: Controlled Dewetting-Dealloying to Fabricate Site-Selective High-activity Nanoporous Au Particles on Highly Ordered TiO$_2$ Nanotube Arrays

*Nhat Truong Nguyen, Marco Altomare, JeongEun Yoo, and Patrik Schmuki\**

N. T. Nguyen, Dr. M. Altomare, J. E. Yoo, Prof. Dr. P. Schmuki
Department of Materials Science and Engineering WW4-LKO, University of Erlangen-Nuremberg, Martensstrasse 7, D-91058 Erlangen, Germany
E-mail: schmuki@ww.uni-erlangen.de

Prof. Dr. P. Schmuki
Department of Chemistry, King Abdulaziz University, Jeddah, Saudi Arabia

Keywords: anodization, TiO$_2$ nanotubes, dewetting, dealloying, photocatalysis

Ever since the groundbreaking work of Fujishima and Honda in 1972,[1] $TiO_2$ has received large attention due to its ability to "photocatalytically" split $H_2O$ into $H_2$ and $O_2$. In the first experiments, Fujishima *et al*. connected a $TiO_2$ film with platinum foil in an electrochemical cell where $H_2$ was generated (cathode) when the $TiO_2$ specimen (anode) was irradiated by solar light. Such a straightforward concept – the illumination of a cheap semiconductor to generate charge carriers that then can be exploited to directly react with water and form $H_2$ – is currently envisaged as a promising and direct path to synthesize $H_2$ as an energy carrier of the future.[2–6]

Besides using an external electrical circuit and an illuminated flat $TiO_2$ electrode (*i.e.*, photo-electrochemistry),[7] another possible approach is the use of $TiO_2$ nanoparticles (NPs) in the form of suspensions for photocatalysis under "open-circuit" conditions (*i.e.*, without applying an external voltage). However, under these conditions, $TiO_2$ is not effective in photo-generating $H_2$ in the absence of a co-catalyst (*i.e.*, charge carrier recombination dominates).[8,9] Most common co-catalysts are noble metals (M) such as Pt, Au or Pd, that, act as electron transfer mediator and $H_2$ recombination catalyst.[9–14] Co-catalyst activity is not only determined by the material properties but also to a large extent by the size and distribution of the noble metal particles.[15] Photocatalytic M@$TiO_2$ systems have therefore been extensively studied in view of optimizing the efficiency of $H_2$ production from water (with or without the usage of sacrificial agents such as methanol or ethanol, among others).

More recently, instead of $TiO_2$ nanoparticles, 1-dimentional (1D) $TiO_2$ nanotubes (NTs) have attracted considerable interest in photocatalysis due to their beneficial charge transport properties and have thus been extensively studied.[4,16] Nevertheless, another feature of the nanotube geometry is that it allows for an exceptionally defined self-ordered arrangement of catalyst particles on the surface. Conventionally, when M@$TiO_2$ systems are fabricated on compact or nanoparticulate $TiO_2$ films, the deposition of metal particles is fairly inhomogeneous,[15] and common techniques provide a relatively low degree of control for optimizing the geometry, size and distribution of the metal particles. In contrast, ordered $TiO_2$ nanotube layers represent an ideally corrugated platform which allows for conducting a well-defined dewetting process of a thin conformal metal layer (*i.e.*, to split it up into highly uniform individual metal particles), and thus achieving a simple but effective control over size and distribution of the co-catalyst NPs. [17] In other words, a highly ordered $TiO_2$ NT platform can be a key benefit in photocatalysis, not only because it provides an optimized semiconductor geometry for photocatalytic reactions, but also as it supplies a hexagonally self-organized array of $TiO_2$ nanocavities that can be exploited for highly-controlled co-catalyst deposition by a simple metal dewetting approach as illustrated in **Scheme 1** and shown in **Figure 1**. The occurrence of thermal dewetting is commonly attributed to the tendency of the metal film to minimize its surface energy on a non-wetting substrate, and consequently to break up forming agglomerates.[18] The morphology of the dewetted particles is strongly influenced by different parameters, which are precisely the metal film thickness, the film homogeneity, the dewetting atmosphere, the substrate morphology and its chemical nature.[12,15,17-25] In general, the locations at which a thin metal film brake up (*i.e.*, "hole formation") are random, but this process can be site-controlled by adopting regularly patterned substrate topographies.

For the dewetting of thin metal films on hexagonally ordered porous oxide substrates, it is reported that specific sites for the concentration of dewetted metal NPs are either at the crown or ground positions of the tube nanostructures (see Scheme 1).[15,25] On the basis of these findings, in the present work we produce catalyst particles – formed under self-organizing conditions – that are dewetted to occupy the ground (Figure 1) and/or crown (**Figure 2**) positions of $TiO_2$ nanotube stumps. We show that it is possible to form well-defined alloyed particles (here Au/Ag) when dewetting sputter deposited Au/Ag double layers, and demonstrate the feasibility to dealloy these particles (so to achieve their porousification) which leads to a further enhancement of their photocatalytic performance.

Scheme 1 outlines the processes used in this work to form a self-ordered TiO$_2$ nanotube platform with various arrangements of Au-based co-catalyst configurations. In a first step we form an array of self-organized NTs by anodizing Ti metal in a hot H$_3$PO$_4$/HF electrolyte, as described in the SI. The selected anodization conditions lead to short aspect ratio tubes (nanotube stumps) with a very high degree of hexagonal order as shown in Figure 1(a) and (b). The individual NTs have an average diameter of ~100 nm and a length of ~200 nm. As shown by the XRD spectra in **Figure 3(a)**, the as-formed TiO$_2$ NTs are amorphous (only the peaks corresponding to the metallic Ti substrate can be seen). In order to form a photocatalytically active structure, the tubes were therefore crystallized into a mixed anatase-rutile TiO$_2$ phase by annealing in air at 450 °C. Except from XRD data, the formation of anatase and rutile polymorphs was also confirmed by HRTEM analysis (**Figure S1**). In these images lattice spacing of 3.5 and 3.3 Å are observed, which correspond well to (101) anatase and (110) rutile planes, respectively.[32,33] Moreover, the XPS data in **Figure 3(b)** show Ti and O peaks with binding energy typical of TiO$_2$.[15]

As illustrated in Scheme 1, these TiO$_2$ nanotube surfaces were then sputter-coated with a thin Au layer (step 2). Figure 1c shows tubes coated with a (nominally) 10 nm thick Au layer. The Au coating in this case is uniformly distributed over the nanotube layers. As visible from the SEM images in Figure 1(a)-(d), a slight coarsening of the tube walls occurs upon Au deposition and consequent narrowing of the tubes inner diameter is also observed. The deposition of Au films with nominal thickness ≥ 10 nm on the tube arrays is evident from the appearance of Au peaks (2θ = 44.6°) in the respective diffractogram (Figure 3(a)). Smaller Au amounts could not be detected by XRD but were apparent in XPS: the deposition of a 1 nm thick Au led to Au XPS peaks at 84.5, 335.6 and 353.2 eV, corresponding to the Au4f, Au4d5 and Au4d3 binding energies (**Figure 3(c)**).[15]

In order to form precursors for alloyed layers (that later were dealloyed), we coated the Au-decorated tubes by a second thin layer of Ag. Figure 1(d) shows an example after sputtering

Ag (20 nm – labeled as Ag20) on the previously deposited Au film (step (3) in Scheme 1). After this, we observed the XRD peak at 2θ = 44.6° to become more intense (Figure 3(a)): this can be ascribed to the overlap of Ag and Au reflections (*i.e.*, Au and Ag have same unit cell and similar lattice constants, and thus contribute both to the same XRD reflections).[34] Further evidence that Au was homogeneously covered by an Ag layer was provided by XPS. We found in fact that after sputtering Ag, the intensity of Au XPS peaks largely decreased, from 17.2 to 3.54 at%. At the same time, the characteristic Ag XPS peaks at 367.9, 573 and 603.4 eV could be detected (**Figure 3(b-d)**).

These regular structures were then used for thermal dewetting (step (4) in Scheme 1).[15] If this done on a flat $TiO_2$ substrate, it leads to the formation of large metal particles of a broad size distribution, randomly spread on the $TiO_2$ surface (**Figure S2**). On the contrary, by using the ordered $TiO_2$ nanotube stump array of Figure 1(a), then self-arranged Au-Ag alloyed particles can be formed in ordered patterns over/within the $TiO_2$ NTs, as illustrated in Figure 1(e), (g) and Figure 2. In other words, this step combines alloying and self-ordering thermal dewetting. The comparison of Figure 1(e) and Figure 2 shows that dewetting can be achieved, leading to a decoration of one particle at the tube bottom ("ground position") and/or to a selective decoration of the tube rims in the "crown position" (step 4 in Scheme 1). We found the key parameters that determine "crown" or "ground" position to be the total loaded amount of metal and the sputtering angle. At sufficiently low loading, shallow sputtering angles lead for instance to a "top-only"–metal coating that after thermal dewetting leads to tubes exclusively decorated with nanoparticles at the "crown" positions.

In every case, the XRD patterns of the dewetted structures show a marked increase of the intensity of Au and Ag peaks (Figure 3(a)) that can be ascribed to a higher degree of crystallinity of the Au-Ag alloyed aggregates obtained by the thermal treatment.

Then, these dewetted Ag-Au particles were dealloyed (step 5 in Scheme 1), aiming at further increasing the performance of the photocatalytic assembly. Here we followed the common concept of dealloying, *i.e.*, that for a certain range of alloy concentrations, a less noble component can be selectively dissolved, leaving behind a porous particle (with higher surface area) of the more noble compound.[29] In our case, the selective removal of Ag was obtained in concentrated nitric acid solutions and, as shown in Figure 1 and Figure 2, and as discussed below in details, indeed nano-sized pores could be detected on the metal NPs after an appropriate treatment. Remarkably, in order to achieve a successful dealloying step, the previous dewetting process is very crucial, *i.e.*, dealloying without dewetting procedure leads only to the formation of non-porous Au particles (**Figure S3**).

First, the XRD pattern of the dealloyed sample shows a decrease of intensity of the metal peaks, this confirming the preferential dissolution of Ag. Accordingly, XPS data reveal a final Ag content of 0.54 at%, compared to 11.45 at% that was measured before dealloying (Figure 3(d) and (e)), proving that nearly all the sputtered Ag could be selectively dissolved. However, we also observed that the Au concentration slightly dropped after the dealloying process (from 3.54 to 2.87 at%), indicating some Au loss during the treatment.

XPS data also support the formation of an Au-Ag alloy by thermal dewetting and the preferential dissolution of Ag by dealloying. The Au and Ag peaks are in fact shifted after dewetting and dealloying, this confirming that both the steps lead to significant change of the chemical surrounding in the metal NPs (Figure 3(c) and (d)). Namely, Au XPS peaks of the nanoporous Au/TiO$_2$ tube arrays show a binding energy of 83.9 eV that was found to be similar to that reported in the literature for dealloyed (porous) Au.[32]

In the case of Figure 2(a) and (b), the TiO$_2$ nanostumps were decorated by Au1-Ag2 (this stands for 1 nm thick Au sputtered film, covered thereafter by 2 nm thick sputtered Ag film). After sputtering, the metal films were dewetted – the SEM images show the particles before and after dealloying.

TEM analysis makes clearer the morphological and crystallographic differences between the metal particles before and after dealloying. Small holes, with size in the range of few nm, were found on each Au particle after the $HNO_3$ treatment (Figure 2(d)), this being an evidence for the occurrence of dealloying (Au-Ag alloy particles show a smooth and hole-free morphology – see Figure 2(c)). In particular, the nanopores occur that are located mainly at the surface of the Au deposits; a common explanation for this is that Au atomic diffusion partially compensates the formation of voids in the Au particles (*i.e.*, inner porosity).[30,31] The lattice spacing measured from HRTEM of a dealloyed metal particle was 1.90 Å (Figure 2(d)), while that of Au (200) is reported in the literature to be 2.03 Å.[33] Such a mismatch in the lattice constant (*ca.* 7%) is attributed to "strain" effects that may arise through dealloying (noteworthy, a mismatch of up to 5% was already reported in the literature for similar systems).[34]

The different catalyst morphologies were then assessed for their photocatalytic performance in hydrogen evolution from ethanol-water mixtures under UV light irradiation. In every case and well in line with some previous reports, we observed for all systems that the amount of photocatalytically evolved $H_2$ linearly increased over the irradiation time (the results discussed below are thus given as the total amount of $H_2$ evolved after a 5h-long photocatalytic run).[12,15,17]

**Figure 4** gives an overview of results for the $H_2$ evolution from Au/Ag-$TiO_2$ systems, where we screened various parameters. Figure 4(a) and (b) illustrate the importance of the amounts of sputtered Ag and Au as such, and their ratio (*i.e.*, Ag/Au ratio), for the photocatalytic performance. The results in Figure 4(a) are obtained from a 5 nm thick Au layer covered with various amounts of Ag that then were dewetted and dealloyed. The results show that a Ag/Au ratio of 2:1 represents an optimized condition for the investigated parameters. Please note that for sputtered films thicker than 15 nm, the $H_2$ evolution efficiency markedly decreased. This can be ascribed to the relatively large amount of co-catalyst (regardless of its type) leading to $TiO_2$ "shading" effects, that is, a smaller photon flux is actually absorbed by the semiconductor.

In order to investigate the influence of the Au loading, we fixed the Ag/Au ratio to 2:1 (Figure 4(b)), and deposited various thicknesses of the metal layers – in line with theory on dewetting mechanisms, the thickness of the metal layer, *i.e.*, the amount, strongly influences dimension and distribution of the dewetted particles (see **Figure S4** and ESI for more details).[18] For the configurations (and loadings) where mixed ground/crown or a full ground position was established, a clearly lower $H_2$ evolution efficiency was obtained compared to only crown position. To shed more light on this aspect, we fabricated sample "5*" by sputtering a (nominally) 5 nm thick Au layer placing the tube arrays in a special configuration, that is, the tube layer was placed parallel to the direction of sputtering, in order to deposit the metal (Au/Ag) film only on crown position. After this, dewetting and dealloying steps followed, that were carried out in otherwise identical conditions. By adopting this special configuration (shallow angle sputtering) the dewetted-dealloyed Au NPs were shown to be placed exclusively at the crown position (see SEM images in Figure 4(b) for comparison). As side-effect of sputtering with shallow angle, the amounts of Au on sample "5*" was of *ca.* 0.23 mg, while those on samples "1" and "2" were of *ca*. 0.16 and 0.30 mg, respectively (Au amounts were measured right after sputtering, for a 2.25 $cm^2$ sample area). As clearly illustrated in Figure 4(b), sample with crown only decoration ("5*") delivered larger amount of $H_2$ compared to the sample "2" (with crown and ground Au decoration). Please note that the former (*i.e.,* "crown" position sample) was decorated with lower amount of Au compared to the latter – these results, highlight the importance of a proper "positioning" of a catalytic particle if one targets the use of a minimal Au amount for achieving a *maximum* photocatalytic performance.

Additional experiments were conducted to explore the effect of the temperature of the dealloying step on the photocatalytic $H_2$ evolution performance. A dealloying process at room temperature (instead of -15 °C, as used for the results in Figure 4(a) and (b)) led to an improvement in $H_2$ evolution, whereas dealloying at 50 °C led to dramatic decrease of photoactivity (Figure 4(c)). In line with the XPS data, this temperature effect seems to be related to the remaining amount of Au after dealloying: in other words, the $HNO_3$ treatment at 50 °C leads to the removal of significant amounts of Au from the tube arrays (loss of co-catalyst), and the $H_2$ evolution is thereafter negatively affected.

To further optimize the efficiency, we prepared photocatalyst films that were dealloyed at room temperature for different times. We observed an *optimum* of duration of the $HNO_3$ treatment around 4 h (Figure 4(d)). In line with previous observations, one may attribute this to an optimal dissolution of Ag without significant release of Au from the metal NPs. Longer dealloying (8 h) led to a five-time smaller $H_2$ evolution, clearly confirming that a right trade-off between metal dissolution/release and Au NP porousification is essential for maximizing the photocatalytic $H_2$ production. Furthermore, the sequence of Au and Ag sputtering was also found to be important for high photocatalytic $H_2$ efficiency (**Figure S5**).

Figure 4(e) gives an overview of the improvement that can be achieved by the individual process steps using the same amount of Au loading on the NT layers. After sputtering only 1 nm of Au, the $H_2$ evolution rate is approximately 10 μL. When simply dewetting this layer into individual Au beads (and therefore following the strategy reported by Yoo),[15] an improvement of nearly 2-fold of $H_2$ generation efficiency is obtained. Using a Ag/Au alloy instead of 1 nm Au, thus introducing dewetting but without a dealloying step, improves the efficiency but only slightly, *i.e.*, up to 25 μL of $H_2$. However, the use of the same alloy and an optimized combination of dewetting/dealloying leads to a nearly 4-times increase in the photocatalyst efficiency or to a 2-times increase of the $H_2$ amount compared to only dewetted layers.

In other words, in comparison with prior work using alloy decoration[14] or simple dewetting,[15] the present work shows how a minimal amount of co-catalyst (*i.e.*, 1 nm of Au) can by an optimized dewetting/dealloying approach be structured to lead to a remarkable photocatalytic enhancement of $H_2$ evolution. A particularly well suited geometrical configuration is the placement of the catalyst at the "crown" position. This may be attributed to the fact that porous gold nanoparticles in "crown" position are more accessible to the liquid phase (diffusion) and the interface contact with reactants is thus maximized, while shadowing is not substantially affecting the performance.[12] In other words, a configuration of Au/TiO$_2$ nanotubes with the noble metal nanoparticles in crown position, which we could not obtain by other previously explored strategies,[14] is seemingly an optimized solution for UV-light driven H$_2$ generation.

Noteworthy, SEM images of the layers after photocatalysis, and results of photo-electrochemical (PEC) measurements show that not only these photocatalysts were stable in the adopted reaction conditions (**Figure S6 and S7(b)** – no Au nanoparticle fall off take place), but could also be successfully used as photo-anodes for PEC water-splitting under simulated solar light irradiation (**Figure S7**).

In summary, the present work shows the fabrication of a photocatalytic platform consisting of anodic TiO$_2$ nanotubes supporting arrays of highly ordered porous Au nanoparticles. The approach is based on using highly ordered TiO$_2$ nanotubes as a morphological guide for an optimized sputtering-dewetting-dealloying sequential approach of a co-catalyst layer. This strategy enables a fine control over the metal nanoparticle size, shape and distribution. The resulting nanoporous Au/TiO$_2$ nanostump assemblies show an enhanced photocatalytic hydrogen production from methanol/water mixtures, ascribed to Au porousification and to the optimized tuning of the co-catalyst amount and positioning over/within the tube arrays.

**Experimental Section**

*Growth of TiO$_2$ nanotubes*: Titanium foils (0.125 mm thickness with 99.6+% purity, Advent Research Materials) were degreased in acetone, ethanol and water for 30 min and then dried in a nitrogen stream. Anodization was conducted in a hot electrolyte composed of hydrofluoric acid (3 M) in phosphoric acid, at 15 V for 2 h and using a DC power supply (VLP 2403 pro, Voltcraft). A platinum foil was used as counter electrode. After anodization, the samples were rinsed with ethanol and dried in a nitrogen stream. Subsequently, the samples were annealed at 450 °C for 30 min in air using a Rapid Thermal Annealer (Jipelec Jetfirst 100 RTA), with a heating and cooling rate of 30 °C min$^{-1}$.

*Nanoparticle decoration*: In order to decorate the TiO$_2$ NTs, plasma sputter deposition (EM SCD500, Leica) was used to deposit different amounts of Au and Ag. The amount of Au and Ag was controlled by measuring their nominal thickness with an automated quartz crystal film-thickness monitor. Thermal dewetting at 400 °C in Ar atmosphere for 30 min was conducted to form alloyed Au-Ag nanoparticles over the TiO$_2$ tubes. For the dealloying process, the dewetted layers were immersed in nitric acid (70 wt%) aqueous solutions (different durations of the treatment and different temperatures were explored).

*Characterization of the structures*: For the morphological characterization of the samples, a field-emission scanning electron microscope (FE-SEM, Hitachi S4800) and a high resolution transmission electron microscope (HR-TEM, Philips CM300) were employed. The chemical composition of the samples was characterized by X-ray photoelectron spectroscopy (XPS, PHI 5600, US). X-ray diffraction (XRD), performed with a X'pert Philips MPD (equipped with a Panalytical X'celerator detector) using graphite monochromized Cu K$_\alpha$ radiation ($\lambda$ =1.54056 Å), was used to analyze the crystallographic properties of the materials.

*Photocatalytic experiments*: The experiments of photocatalytic H$_2$ production were conducted by irradiating with UV light (325 nm, 60 mW cm$^{-2}$, HeCd laser, Kimmon, Japan) samples immersed in aqueous solutions of ethanol (20 vol%) for 5 h. The experiments were run in a quartz tube photocatalytic cell, sealed up with a rubber septum (to accumulate the evolved gas) from which samples (200 μL) were withdrawn and analyzed by gas chromatography (GCMS-QO2010SE, Shimadzu) to determine the amount of photocatalytically produced H$_2$. The GC was equipped with a thermal conductivity detector (TCD), a Restek micropacked Shin Carbon ST column (2 m x 0.53 mm). GC measurements were carried out at a temperature of the oven of 45 °C (isothermal conditions), with the temperature of the injector set up at 280 °C and that of the TCD fixed at 260 °C. The flow rate of the carrier gas, *i.e.*, argon, was 14.3 mL min$^{-1}$.

*Photoelectrochemical experiments*: The photoelectrochemical water-splitting experiments were performed in KOH aqueous solution (1 M) with a three-electrode configuration, consisting of a TiO$_2$ NTs layer photoanode used as the working electrode, a saturated Ag/AgCl electrode as the reference and a platinum foil as the counter electrode. Photocurrent density vs. potential measurements were carried out under AM 1.5 simulated solar light irradiation (300 W Xe, 100 mW cm$^{-2}$, illuminated area = 0.38 cm$^2$) with a scanning potentiostat (Jaissle IMP 88 PC, scan rate of 1 mV s$^{-1}$). Photocurrent density vs. time measurements were conducted at a constant potential of 0.5 V (vs. Ag/AgCl) under simulated solar light irradiation. The light intensity was measured prior to the experiments using a calibrated Si photodiode. Photocurrent spectra were recorded at a constant potential of 0.5 V (vs. Ag/AgCl) with a potentiostat (Jaissle IMP83 PC-T-BC) in Na$_2$SO$_4$ aqueous solutions (0.1 M) under 200-800 nm light source (Oriel 6365 150 W Xe-lamp), while the wavelength was varied using a motor driven monochromator (Oriel Cornerstone 130 1/8 m). Incident photon to energy conversion (IPCE) values were calculated using IPCE %= (1240*$i_{ph}$)/($\lambda$*$I_{light}$), where $i_{ph}$ is the photocurrent density (mA cm$^{-2}$), $\lambda$ is the incident light wavelength (nm), and $I_{light}$ is the intensity of the light source at each wavelength (mW cm$^{-2}$).

**Acknowledgements**
The authors would like to acknowledge the ERC, the DFG and the DFG cluster of excellence, EAM for financial support as well as H. Hildebrand for valuable technical help. Dr. G. Loget is acknowledged for valuable discussions.

**References**

[1]     A. Fujishima, K. Honda, *Nature* **1972**, *238*, 37.

[2]     M. Ni, M. K. H. Leung, D. Y. C. Leung, K. Sumathy, *Renew. Sustain. Energy Rev.* **2007**, *11*, 401.

[3]     X. Chen, S. S. Mao, *Chem. Rev.* **2007**, *107*, 2891.

[4]     K. Lee, A. Mazare, P. Schmuki, *Chem. Rev.* **2014**, *114*, 9385.

[5]     X. Wang, Z. Li, J. Shi, Y. Yu, *Chem. Rev.* **2014**, *114*, 9346.

[6]     A. Fujishima, X. Zhang, D. Tryk, *Surf. Sci. Rep.* **2008**, *63*, 515.

[7]     G. Wang, H. Wang, Y. Ling, Y. Tang, X. Yang, R. C. Fitzmorris, C. Wang, J. Z. Zhang, Y. Li, *Nano Lett.* **2011**, *11*, 3026.

[8]     N. Liu, C. Schneider, D. Freitag, M. Hartmann, U. Venkatesan, J. Müller, E. Spiecker, P. Schmuki, *Nano Lett.* **2014**, *14*, 3309.

[9]     N. Liu, C. Schneider, D. Freitag, U. Venkatesan, V. R. R. Marthala, M. Hartmann, B. Winter, E. Spiecker, A. Osvet, E. M. Zolnhofer, K. Meyer, T. Nakajima, X. Zhou, P. Schmuki, *Angew. Chem. Int. Ed.* **2014**, *53*, 14201.

[10]    M. Anpo, *J. Catal.* **200**3, *216*, 505.

[11]    M. Murdoch, G. I. N. Waterhouse, M. A. Nadeem, J. B. Metson, M. A. Keane, R. F. Howe, J. Llorca, H. Idriss, *Nat. Chem.* **2011**, *3*, 489.

[12]    N. T. Nguyen, J. Yoo, M. Altomare, P. Schmuki, *Chem. Commun.* **2014**, *50*, 9653.

[13]    R. Su, R. Tiruvalam, A. J. Logsdail, Q. He, C. A. Downing, M. T. Jensen, N. Dimitratos, L. Kesavan, P. P. Wells, R. Bechstein, H. H. Jensen, S. Wendt, C. R. A. Catlow, C. J. Kiely, G. J. Hutchings, F. Besenbacher, *ACS Nano* **2014**, *8*, 3490.

[14]    K. Lee, R. Hahn, M. Altomare, E. Selli, P. Schmuki, *Adv. Mater.* **2013**, *25*, 6133.

[15]    J. E. Yoo, K. Lee, M. Altomare, E. Selli, P. Schmuki, Angew. *Chem. Int. Ed.* **2013**, *52*, 7514.

[16]    P. Roy, S. Berger, P. Schmuki, *Angew. Chem. Int. Ed.* **2011**, *50*, 2904.

[17]    J. Yoo, K. Lee, P. Schmuki, *Electrochem. Commun.* **2013**, *34*, 351.

[18]    C. V. Thompson, *Annu. Rev. Mater. Res.* **2012**, *42*, 399.

[19]    A. L. Giermann, C. V. Thompson, *Appl. Phys. Lett.* **2005**, *86*, 121903.

[20]    A. Geissler, M. He, J.M. Benoit, P. Petit, *J. Phys. Chem. C* **2010**, *114*, 89.

[21]    Y. J. Oh, C. A. Ross, Y. S. Jung, Y. Wang, C. V. Thompson, *Small* **2009**, *5*, 860.

[22]    S. Yang, F. Xu, S. Ostendorp, G. Wilde, H. Zhao, Y. Lei, *Adv. Funct. Mater.* **2011**, *21*, 2446.

[23]    T. Karakouz, D. Holder, M. Goomanovsky, A. Vaskevich, I. Rubinstein, *Chem. Mater.* **2009**, *21*, 5875.

[24]    G. Y. Kim, C. V. Thompson, *Acta Mater.* **2015**, *84*, 190.

[25]    S. Hong, T. Kang, D. Choi, Y. Choi, L. P. Lee, *ACS Nano* **2012**, *6*, 5803.

[26]    M. Liu, L. Piao, L. Zhao, S. Ju, Z. Yan, T. He, C. Zhou, W. Wang, *Chem. Commun.* **2010**, *46*, 1664.

[27]    B. Liu, E. S. Aydil, *J. Am. Chem. Soc.* **2009**, *131*, 3985.

[28]    J.-H. Liu, A.-Q. Wang, Y.-S. Chi, H.-P. Lin, C.-Y. Mou, *J. Phys. Chem. B* **2005**, *109*, 40.

[29]    A. J. Forty, *Nature* **1979**, *282*, 597.

[30]    L. H. Qian, M. W. Chena, *Appl. Phys. Lett.* **2007**, *91*, 083105.

[31]    J. Erlebacher, M. J. Aziz, A. Karma, N. Dimitrov, K. Sieradzki, *Nature* **2001**, *410*, 450.

[32]    C. Xu, J. Su, X. Xu, P. Liu, H. Zhao, F. Tian, Y. Ding, *J. Am. Chem. Soc.* **2007**, *129*, 42.

[33]    H. Ohnishi, Y. Kondo, K. Takayanagi, *Nature* **1998**, *395*, 780.

[34]    T. Fujita, P. Guan, K. McKenna, X. Lang, A. Hirata, L. Zhang, T. Tokunaga, S. Arai, Y. Yamamoto, N. Tanaka, Y. Ishikawa, N. Asao, Y. Yamamoto, J. Erlebacher, M. Chen, *Nat. Mater.* **2012**, *11*, 775.

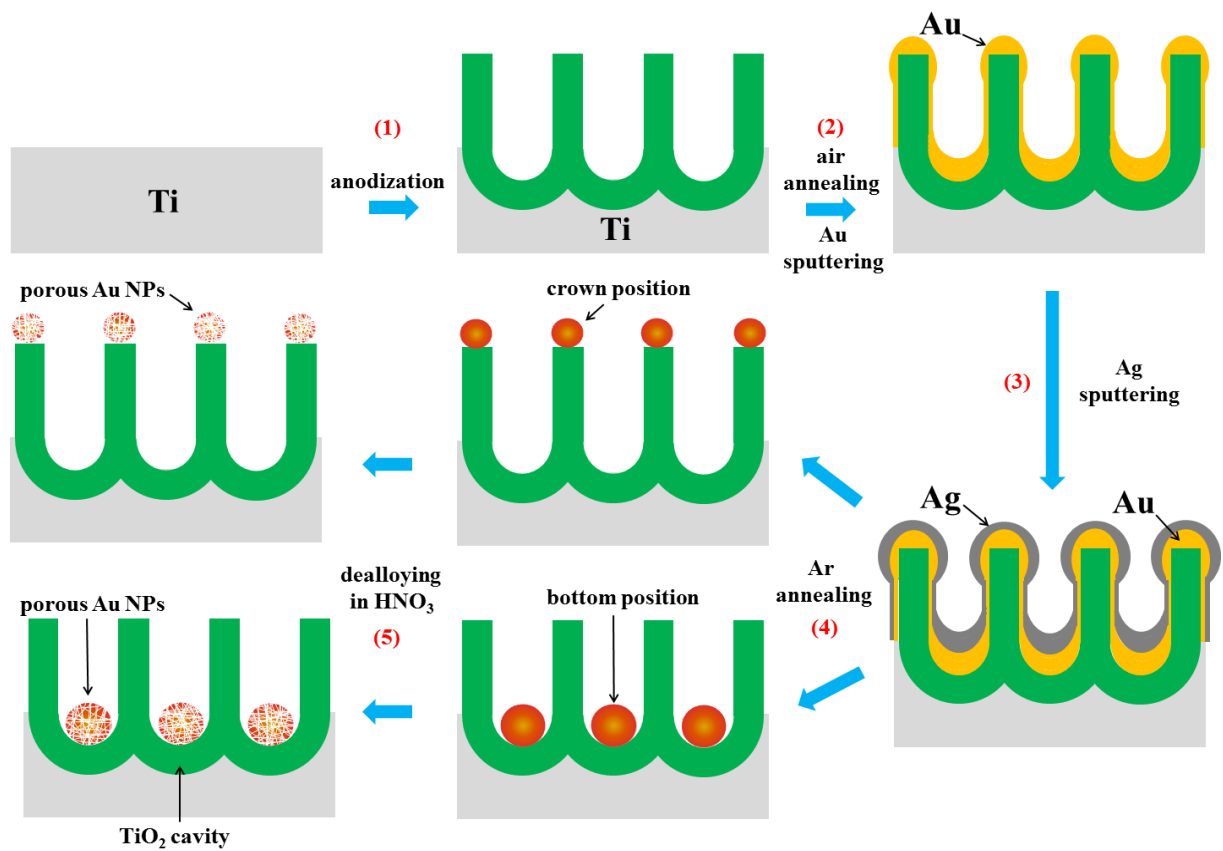

**Scheme 1.** Formation of nanoporous Au over/within anodic $TiO_2$ nanotubes: (1) tube array fabrication by anodization in $H_3PO_4$/HF, (2) air annealing and Au sputtering, (3) Ag sputtering, (4) formation of Au-Ag alloyed NP array by thermal dewetting in Ar and (5) formation of porous Au NP array by dealloying in $HNO_3$.

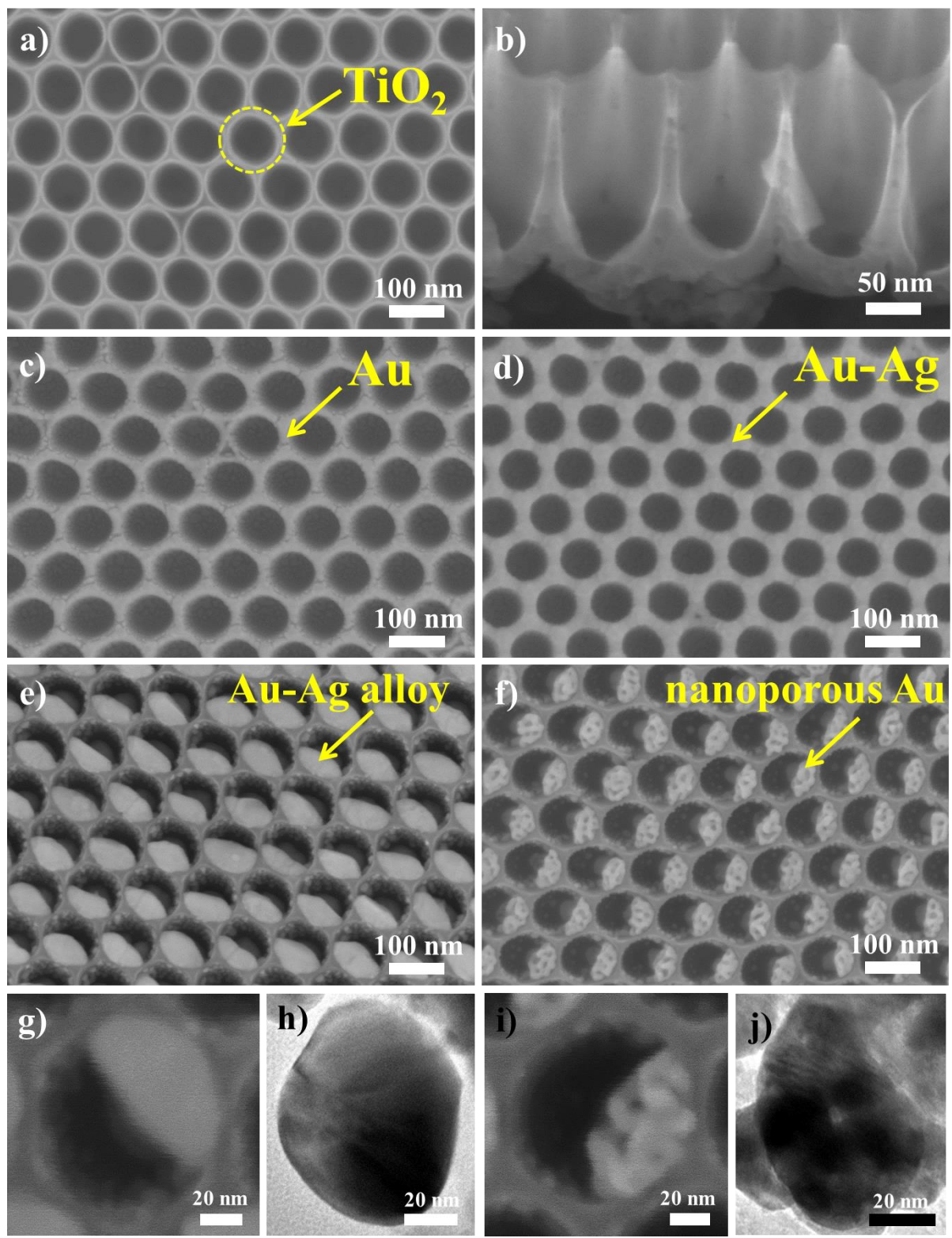

**Figure 1** SEM images of: (a,b) as form $TiO_2$ NTs, (c) NTs after 10 nm Au sputtering, (d) NTs after 10 nm Au and 20 nm Ag sputtering, (e,g) Au-Ag alloyed NP after thermal dewetting in Ar and (f,i) porous Au NP after dealloying in $HNO_3$. TEM images of: (h) Au-Ag alloyed NP and (j) porous Au NP after dealloying.

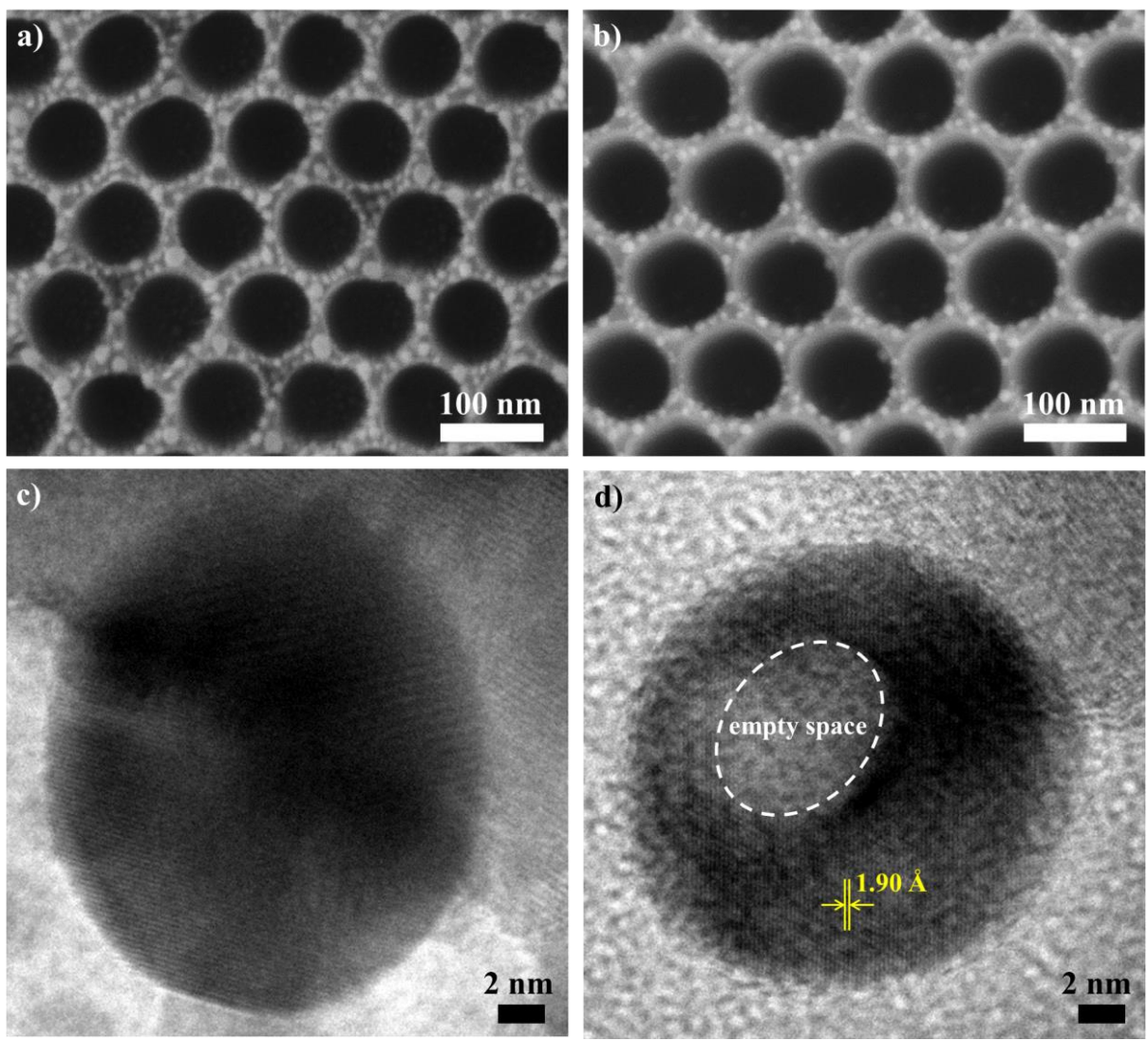

**Figure 2** (a, b) SEM and (c, d) TEM images of (a, c) TiO$_2$ NTs decorated with Au-Ag alloyed-dewetted NPs (1 nm Au – 2 nm Ag), and (b, d) porous Au NPs after dealloying.

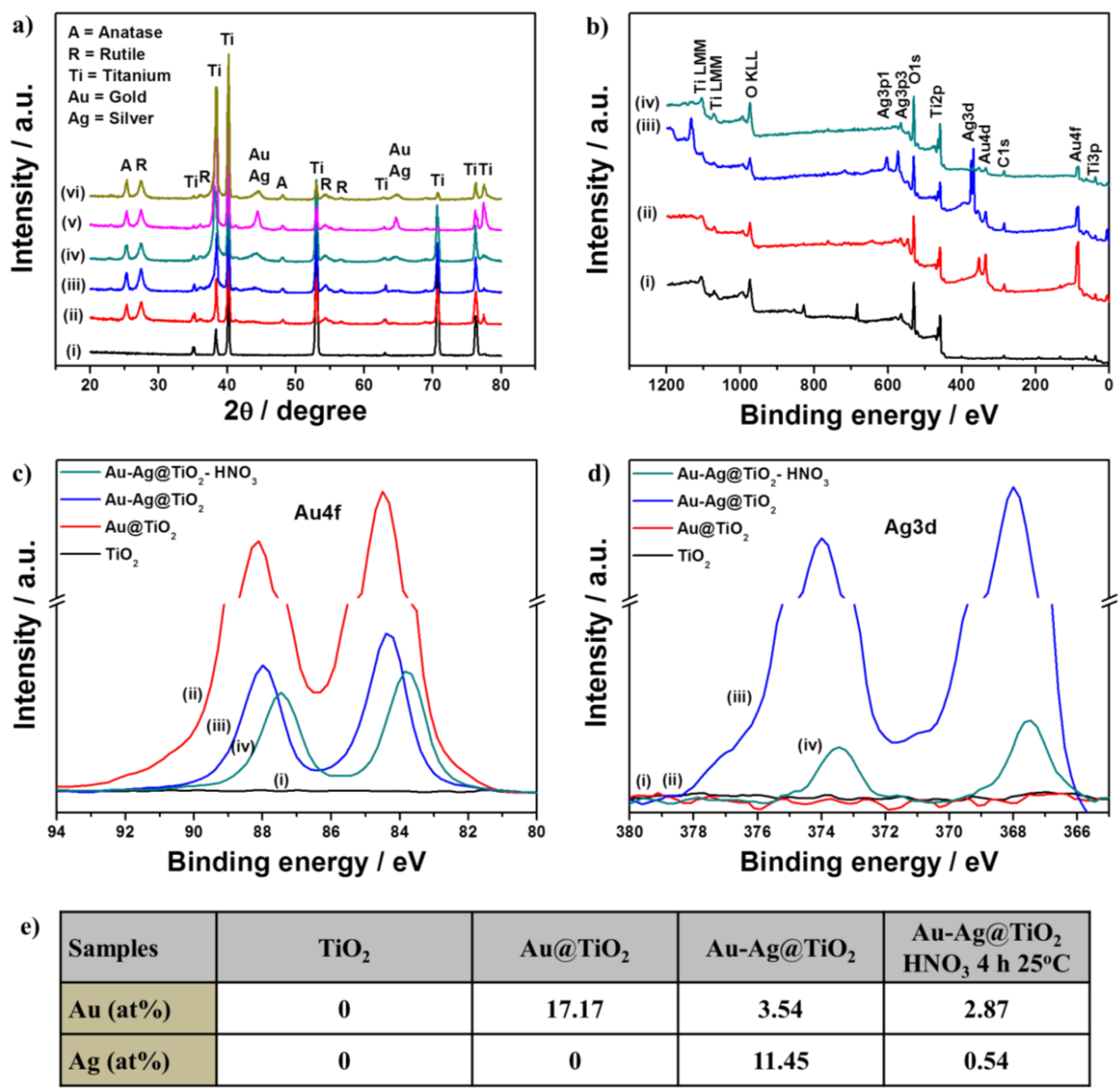

**Figure 3** (a) XRD data of: (i) as-formed $TiO_2$ NTs, (ii) annealed $TiO_2$ NTs, (iii) Au10@$TiO_2$ NTs, (iv) Au10Ag20@$TiO_2$ NTs, (v) Au10Ag20@$TiO_2$ NTs after thermal dewetting and (vi) Au10Ag20@$TiO_2$ NTs after thermal dewetting and dealloying in $HNO_3$. (b) XPS survey. (c) and (d) XPS high-resolution patterns of: (i) as-formed $TiO_2$ NTs, (ii) Au1@$TiO_2$ NTs, (iii) Au1Ag2@$TiO_2$ NTs and (iv) Au1Ag2@$TiO_2$ NTs after thermal dewetting and dealloying in $HNO_3$ for 4 h at 25 °C. (e) Au and Ag concentration determined by XPS data in (c, d) for the tube arrays at different step of their functionalization.

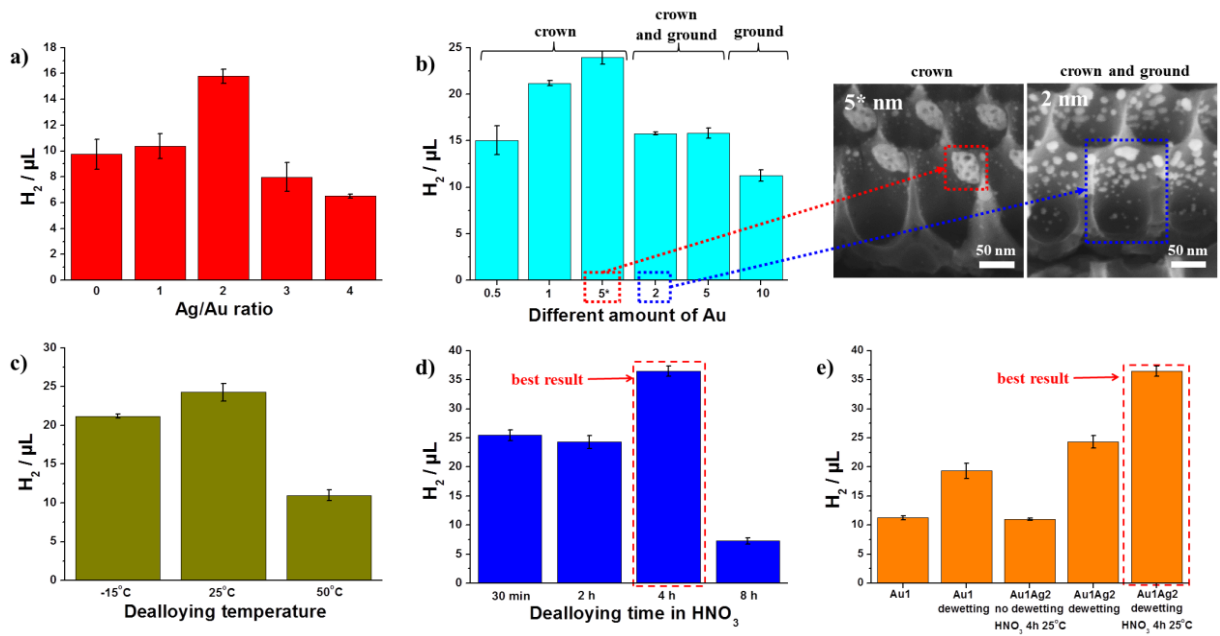

**Figure 4** Photocatalytic $H_2$ evolution measured for: (a) Different Ag/Au ratio, all samples were prepared by sputtering 5 nm Au and different amount of Ag, followed by dewetting and dealloying in $HNO_3$ for 2 h at -15 °C. (b) Different amount of Au (nm) with constant Ag/Au ratio of 2:1, all samples were dewetted and dealloyed in $HNO_3$ for 2 h at -15 °C. (c) Au1Ag2@$TiO_2$ NTs after dealloying in $HNO_3$ for 2 h at different temperatures. (d) Au1Ag2@$TiO_2$ NTs after dealloying in $HNO_3$ at 25 °C for different times. (e) Different systems all prepared by depositing a 1 nm nominally thick layer of Au (the plot highlights that an optimized combination of dewetting/dealloying leads to a nearly 4-time increase of the $H_2$ generation efficiency). SEM images are relative to the samples labelled in (b) as "5*" and "2". All the photocatalytic runs lasted 5 h and were performed under UV light irradiation (HeCd laser, 325 nm, 60 mW cm$^{-2}$).

**Table of contents**

**Overlaid self-ordering processes:** we form anodic self-organized TiO$_2$ nanostumps and exploit them for self-ordering dewetting of Au-Ag sputtered films. This forms ordered particle configurations at tube top (crown position) or bottom (ground position). By dealloying we then form, from minimal amount of noble metal, porous Au nanoparticles that when in crown position allow for a drastically improved photocatalytic H$_2$ production compared with nanoparticles produced by conventional dewetting processes.

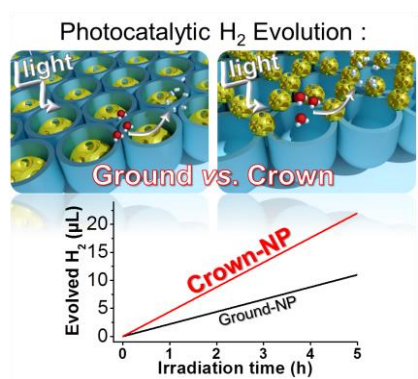

# Supporting Information

**Efficient Photocatalytic $H_2$ Evolution: Controlled Dewetting-Dealloying to Fabricate Site-Selective High-activity Nanoporous Au Particles on Highly Ordered $TiO_2$ Nanotube Arrays**

*Nhat Truong Nguyen, Marco Altomare, JeongEun Yoo, and Patrik Schmuki\**

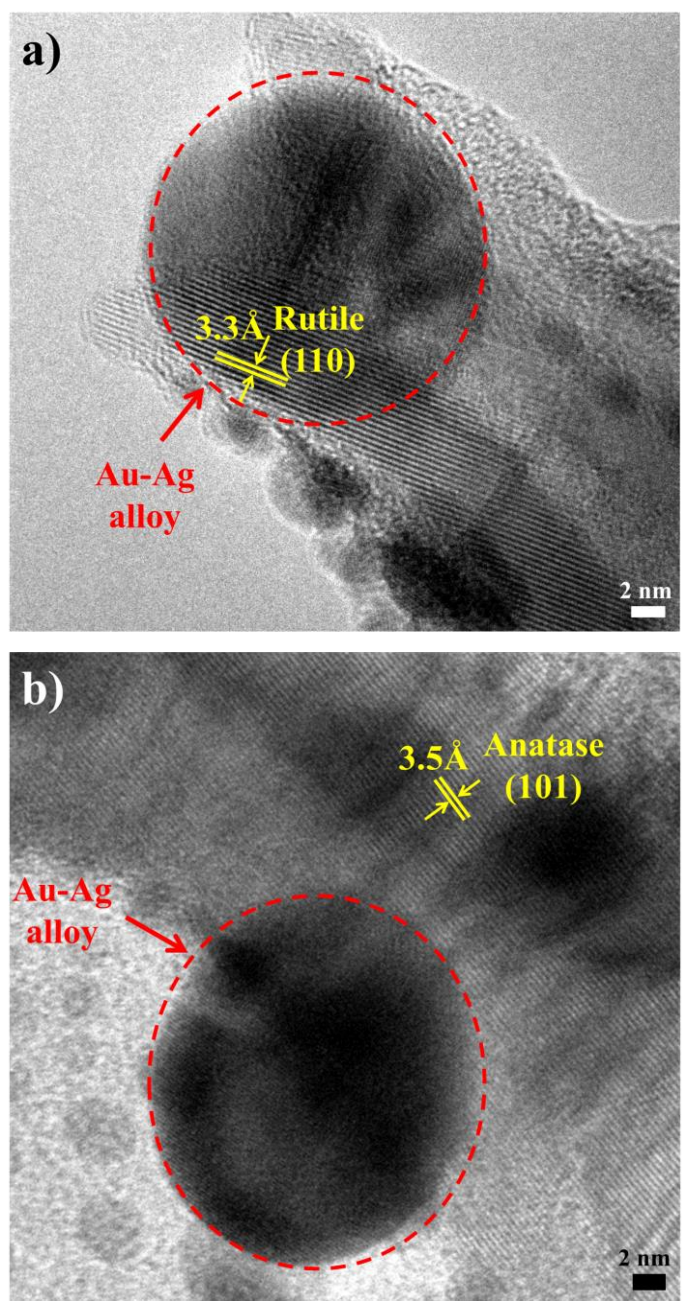

**Figure S1** HRTEM images of TiO$_2$ nanotubes after 1 nm Au - 2 nm Ag sputtering followed by thermal dewetting in Ar, showing (a) rutile (110) and (b) anatase (101) TiO$_2$ crystallographic planes.

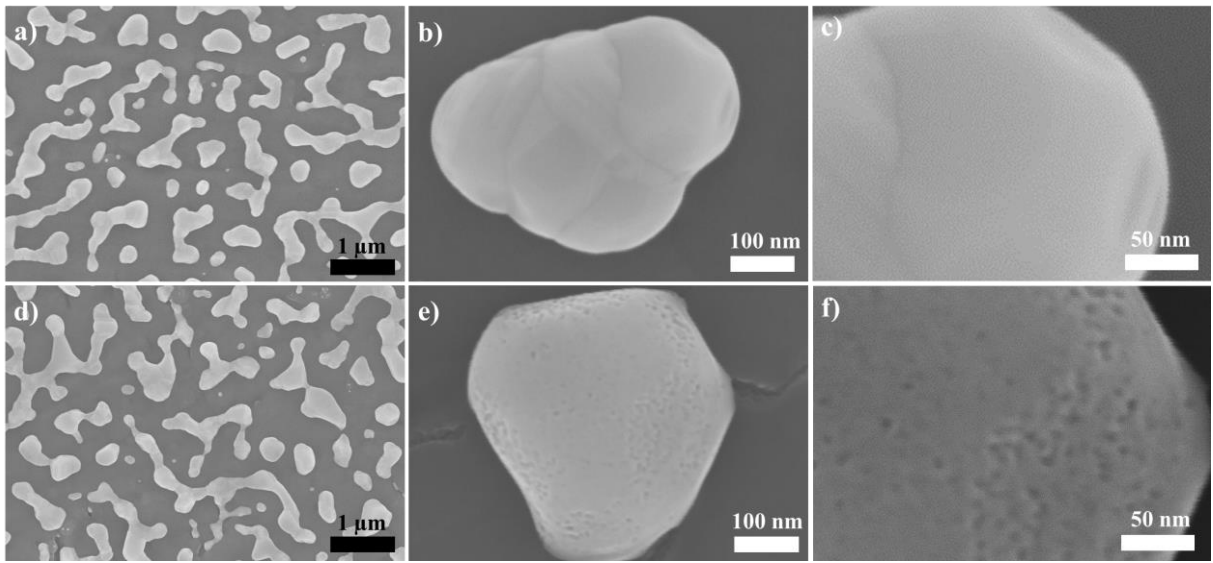

**Figure S2** SEM images (different magnifications) of a TiO$_2$ compact layer decorated by a), b), c) dewetting a Au 10 nm – Ag 20 nm film, and d), e), f) dewetting-dealloying a Au 10 nm – Ag 20 nm film in HNO$_3$ at -15 °C for 2 h. The compact TiO$_2$ layers (*i.e.*, the substrates) were formed by anodization of titanium foils in 1 M H$_2$SO$_4$ at 20 V for 15 min.

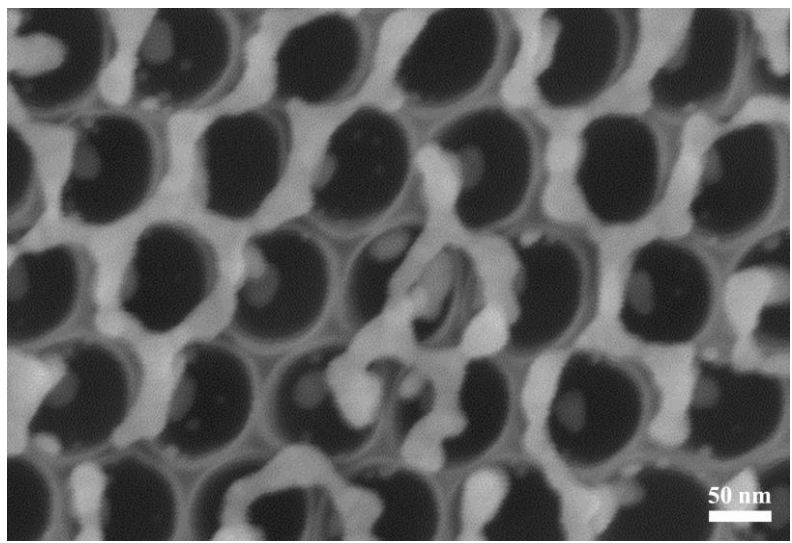

**Figure S3** SEM image of TiO$_2$ nanotubes decorated by dealloying a Au 10 nm – Ag 20 nm film (without dewetting process) in HNO$_3$ at 25 °C for 4 h.

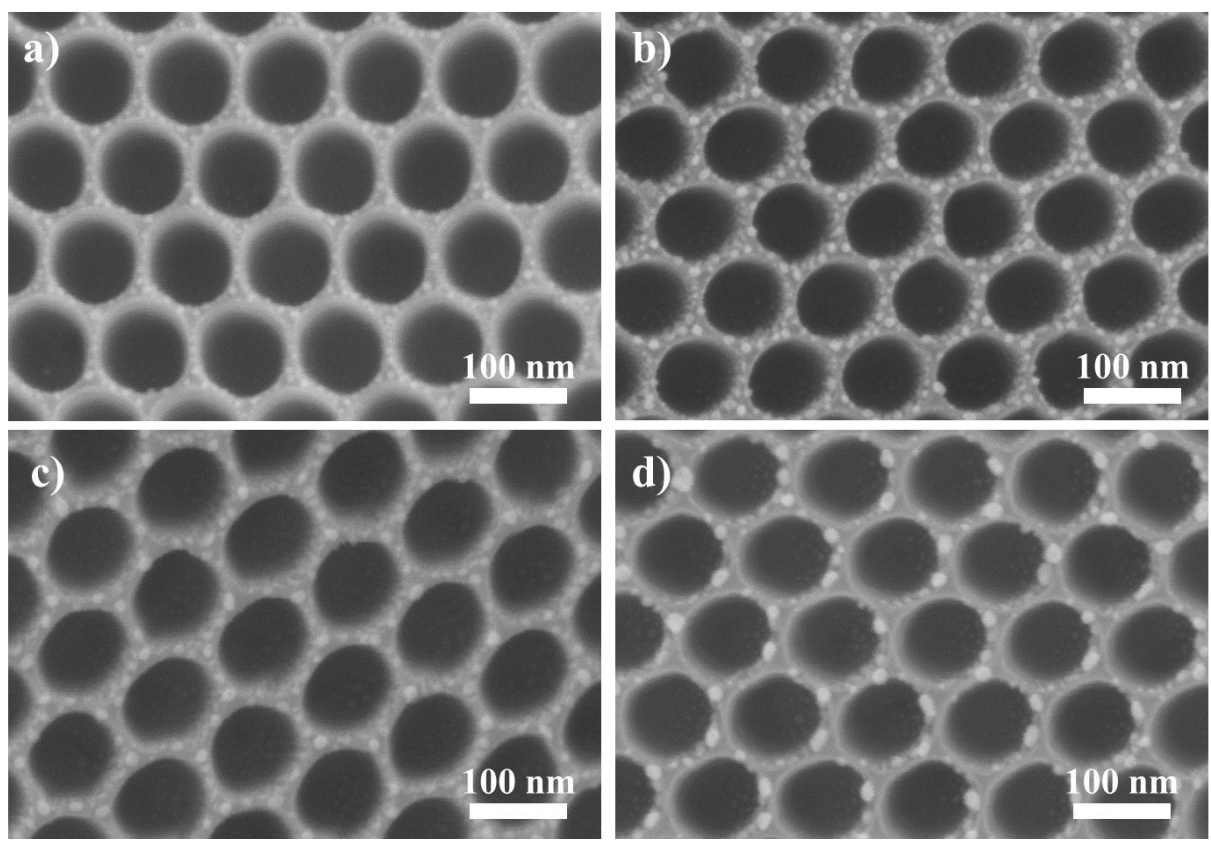

**Figure S4** SEM images of TiO$_2$ nanotubes decorated by dewetting-dealloying film of a) Au 0.5 nm – Ag 1 nm, b) Au 1 nm – Ag 2 nm, c) Ag 2 nm – Au 1 nm, and d) Au 2 nm – Ag 4 nm. All samples were dealloyed in HNO$_3$ at -15 °C for 2 h.

Moreover, well in line with theory on the dewetting mechanism,[1] we found that the mean size of the Pt nanoparticles after thermal dewetting strictly depended on the thickness of the initially deposited Pt layer

[1] C. V. Thompson, *Annu. Rev. Mater. Res.* **2012**, *42*, 399.

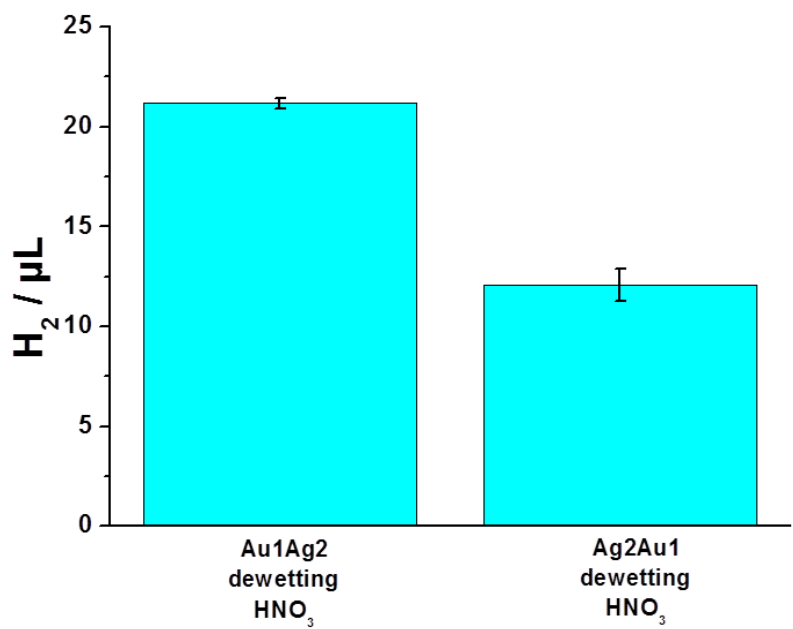

**Figure S5** Photocatalytic H$_2$ evolution measured with TiO$_2$ nanotubes decorated by dewetting-dealloying approach, using opposite sequences for the metal film sputtering.

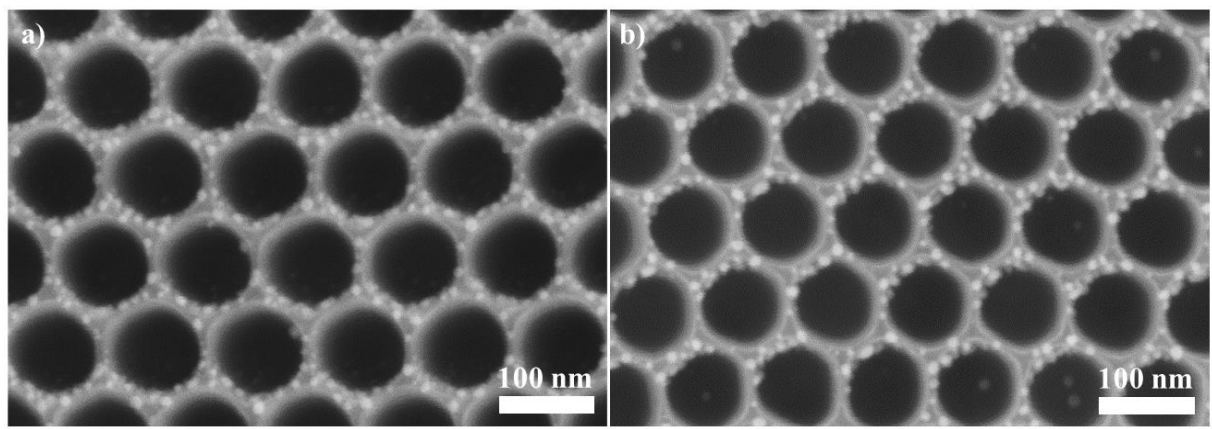

**Figure S6** SEM images a) before and b) after 5 h-long photocatalysis for TiO$_2$ nanotubes decorated by dewetting-dealloying Au 1 nm – Ag 2 nm. Dealloying was performed in HNO$_3$ at 25 °C for 4 h.

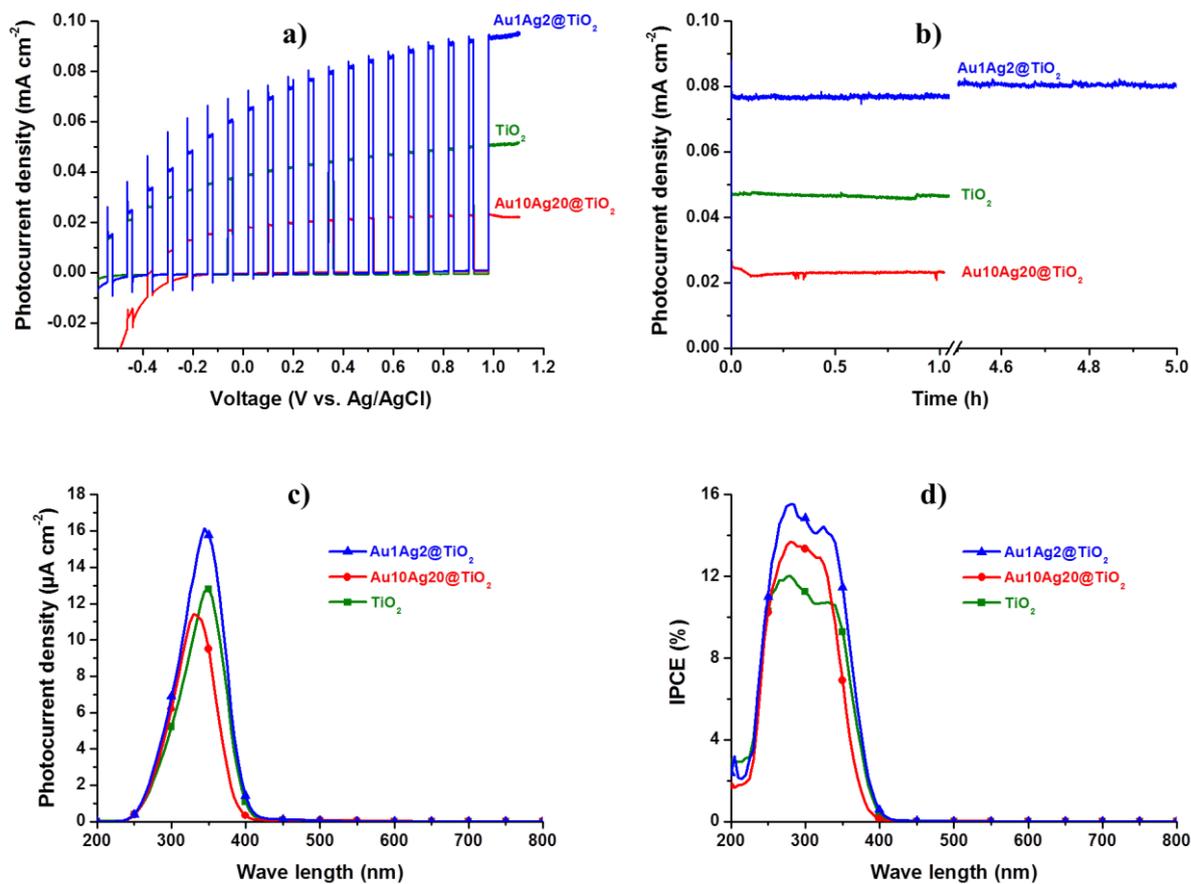

**Figure S7** a) Photocurrent density vs. potential curves under chopped solar light illumination and b) photocurrent density vs. time under solar light illumination of $TiO_2$ NTs and $TiO_2$ NTs decorated with Au-Ag alloyed-dewetted NPs (1 nm Au – 2 nm Ag and 10 nm Au – 20 nm Ag). c) Photocurrent spectra and d) IPCE measurements of $TiO_2$ NTs and $TiO_2$ NTs decorated with Au-Ag alloyed-dewetted NPs (1 nm Au – 2 nm Ag and 10 nm Au – 20 nm Ag).

26